\documentclass[a4paper]{eccomas_2022}
\usepackage{graphicx}
\usepackage{amsmath}
\usepackage{amsfonts}
\usepackage{amssymb}
\usepackage{graphicx}
\usepackage{amsmath}
\usepackage{amsfonts}
\usepackage{amssymb}
\usepackage{subcaption}
\usepackage{todonotes}
\usepackage[backend=biber,bibencoding=utf8]{biblatex}
\usepackage{wrapfig}
\usepackage{url}

\title{PROBABILISTIC MODELING OF LCF FAILURE TIMES USING AN EPIDEMIOLOGICAL CRACK PERCOLATION MODEL}

\author{$^{*}$M. HARDER$^{1}$, P. LION$^{2}$, L. M\"ADE$^{3}$, T. BECK$^{2}$ AND H. GOTTSCHALK$^{1}$}

\heading{M. Harder, P. Lion, L. M\"ade, T. Beck and H. Gottschalk}

\address{$^{1}$ IMACM, Bergische Universit\"at Wuppertal, Gau{\ss}stra{\ss}e 20 D-42119 Wuppertal, \{mharder,hgottsch\}@uni-wuppertal.de
\and
$^{2}$ TU Kaiserslautern, Postfach 3049 D-67653 Kaiserslautern, \{plion,beck\}@mv.uni-kl.de
\and
$^{3}$ Siemens Energy Global GmbH \& Co. KG
Enterprise data and advanced analytics, D-10553 Berlin, lucas.maede@siemens-energy.com
}
\keywords{Probabilistic modeling of low cycle fatigue $\bullet$ anisotropic microstructure $\bullet$ epidemiological modeling of crack propagation }

\abstract{The analysis of standardized low cycle fatigue (LCF) experiments shows that the failure times widely scatter.
Furthermore, mechanical components often fail before the deterministic failure time is
reached. A possibility to overcome these problems is to consider probabilistic failure
times.
Our approach for probabilistic life prediction is based on the microstructure of the metal.
Since we focus on nickel-base alloys we consider a coarse grained microstructure, with random oriented FCC grains. This leads to random distributed Schmid factors and different
anisotropic stress in each grain. To gain crack initiation times, we use Coffin-Manson-
Basquin and Ramberg-Osgood equation on stresses corrected with probabilistic Schmid
factors. \\
Using these single grain crack initiation times, we have developed an epidemiological crack
growth model over multiple grains. In this mesoscopic crack percolation model, cracked
grains induce a stress increase in neighboring grains. This stress increase is realized using
a machine learning model trained on data generated from finite element simulations. The
resulting crack clusters are evaluated with a failure criterion based on a multimodal stress
intensity factor. 
From the generated failure times, we calculate surface dependent hazard
rates using a Monte Carlo framework.
We compare the obtained failure time distributions to data from LCF experiments and find good coincidence of predicted and measured scatter bands.}

\begin{document}

\section{INTRODUCTION}
Fatigue is the process under which metals degrade under cyclic load and thereby fail after a given number of load cycles with load amplitudes well below the ultimate tensile stress (UTS) \cite{RoeslerHardersBaeker_2012}. Here, UTS is defined as the load that leads to immediate failure. Low cycle fatigue (LCF) stands for cyclic fatigue at load amplitudes that lead to failure within a few to several thousands of load cycles. In such processes, plastic deformation of the material play a pivotal r\^ole.  

The microstructure of a conventionally cast metal consists of microscopic grains, in which atoms are organized in crystallographic lattices. In the lattice structure, one dimensional lattice defects can travel under the influence of cyclic loads. For such traveling dislocations, the activation energy required is lowest along certain slip systems, i.e. planes of densest packing and directions on the planes in the crystal's lattice structure \cite{Gottstein_2013}. This points to a decisive influence of the local orientation of grain structures. As the latter acquire random orientation during the solidification process in a conventional cast, fatigue processes that depend on grain orientation become random processes themselves \cite{EngelBeck_2018,EngelMaede_2019,EngelLCF8_2017,gottschalk2015probabilistic}.     

This is especially true for Nickel (Ni-) based superalloys as are applied in high temperature applications like gas turbine design \cite{SjobergCaballero_2004,WahlHarris_2001}. Superalloys often come with a comparatively coarse grain structure, which prevents homogenization effects at scales which are relevant for the design. Furthermore, the elasticity constants of Ni-based superalloys are highly anisotropic \cite{RoeslerHardersBaeker_2012} and thereby enhance the effect of a statistical scatter of LCF lifetime, as the local elasticity constants contribute to the calculation of intragranular stress state.

That the LCF lifetime is subject to a high level of statistical scatter has been observed for a long time. In \cite{SchmitzSeibel_2013,SchmitzPhD_2014}, some of the authors proposed an empirical local model for low cycle fatigue on the basis of a Poisson process model. This line of research has been applied to gas turbine design \cite{SchmitzRollmann_2013,SchmitzPhD_2014}, shape optimization \cite{bittner2016optimal,GottschSchmitz_2012,gottschalk2018adjoint,gottschalk2019shape,gottschalk2021analytical} and tolerance design \cite{liefke2020towards}. Further developments include notch support factors into the model \cite{MaedeSchmitzASME_2017,MaedeSchmitz_2018}, see also  \cite{babuvska2019spatial,liao2018computational,liao2020probabilistic,liao2020recent,zhu2018evaluation} for related approaches. While these empirical models have been successfully tested experimentally, they lack a justification based on the underlying micro-mechanical phenomena.

In \cite{GottschalkSchmitz_2015} some of the authors proposed a model based on the Schmid factors that model the maximum shear stress on any crystallographic slip system depending on its relative orientation to the stress tensor. This model has been subsequently extended considering also the local anisotropy of the metal's elastic constants \cite{EngelLCF8_2017,EngelBeck_2018,EngelMaede_2019,EngelOhneseit_2019}. A model purely based on single grain probability of failure (PoF), defining a metallic structure as failing when the first grain is cracked, leads to a major underestimation of the statistically observed scatter. The mathematical reason for the failure is the narrowing in effect of extremal value distributions \cite{de2006extreme} while the physical reason can be associated with the experimental definition of a technically relevant mesoscopic crack that extends over several microscopic grains.  

Crack percolation modeling \cite{bollobas2006percolation} thereby is a constitutive part of any successful physics based model for the scatter of the number of cycles to (mesoscopic) crack initiation. The model presented in \cite{MochPhD_2018} includes multi grain crack percolation and in \cite{maede_PhD} this model is refined and tested extensively against experimental data. These models however do not model any interaction between a crack already formed in one grain which will influence the crack formation in neighboring grains as loads are locally re-distributed from the cracked grain to the adjacent grains.

In this work, we thus undertake an effort to model such interaction effects in a physics based probabilistic model of LCF. To this aim, we perform finite element (FEM) simulations \cite{ern2004theory} modeling the influence of a intragranular crack in a crystal with random orientation on the shear stress on randomly oriented slip systems on the other side of the grain boundary. In this way, we extract data on the relative increase or decrease of these stresses. Based on this data, we train a gradient boosting machine \cite{hastie2009elements}  that is able to predict these stress deviations with respect to the stress state in the same grain with an non cracked neighbor grain on the basis of the in total six Euler angles of both grains.  

The probabilistic grain-micostructure model we employ consists of random Voronoi tesselations \cite{torquato2002random}, where we equip each Voronoi cell with a crystal orientation. Our model starts with the single grain crack formation model up to the time, where the first grain is cracked. Thereafter the stresses of neighboring grains are corrected according to our machine learning model, which (almost always) leads to an accelerated consumption of the remaining life. This adds an epidemiological aspect to the model. In this way, clusters of cracked grains spread more rapidly as compared to the percolation model to the size of a technically relevant crack, which leads to a slightly reduced scatter bands. Here the notion of technically relevant cracks, following \cite{GrossSeelig_2001}, are based on multiaxial critical stress concentration factors. 

Finally, we calibrate this epidemiological model with data from isothermal LCF experiments, which shows a reasonable prediction of the scatter band after an adequate fitting of the model's parameters. 
\begin{figure}[h]
    \centering
    \includegraphics[width=\textwidth]{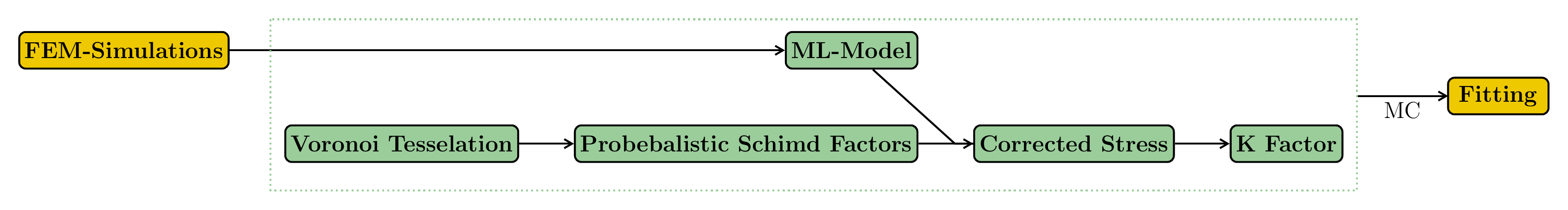}
    \caption{Overview of the Modeling Process. The blue steps are in am Monte Carlo loop.}

\end{figure}

Our paper is organized as follows: In Section \ref{sec:Schmid} we recapitulate the model of random Schmid factors. The following Section \ref{sec:FEM} presents the anisotropic FEM-simulations of two adjacent grains, one of which is cracked. This section also discusses the training of the deep neural network to predict Schmid factors based on the  orientation of both grains. Section \ref{sec:Percolation} then describes our epidemiological percolation model, which is fitted and compared to experimental data in Section \ref{sec:Fitting}. We end with some concluding remarks in Section \ref{sec:Conclusion}.  

\section{PROBABILISTIC SCHMID FACTORS}
\label{sec:Schmid}

Due to the crystalline structure of metals, the elastic properties of individual grains are directional. The orientation of the grains influences both the anisotopic
stress and the orientation of the
slip systems. In a non-textured polycrystal, 
the orientation of the grains is randomly distributed. The orientation of a grain can be described by three Euler angles in the Bungee convention $(\varphi_1, \theta, \varphi_2)$.  Each angle represents a rotation around the default crystal coordinate axes. Therefore the rotation can be represented as a $3 \times 3$-matrix 
\vskip-.6cm
\begin{eqnarray}\label{rot_m}
U(\varphi_1, \theta, \varphi_2) = \begin{pmatrix}
c(\varphi_1) c(\varphi_2) - s(\varphi_1) c(\theta) s(\varphi_2) & -c(\varphi_1) s(\varphi_2) - s(\varphi_1) c(\theta) c(\varphi_2) & s(\theta) s(\varphi_1)\\
s(\varphi_1) c(\theta) + c(\varphi_1) c(\theta) s(\varphi_2) & -s(\varphi_1) s(\theta) + c(\varphi_1) c(\varphi_2) & -s(\theta) s(\varphi_1) \\
s(\theta) s(\varphi2)  & s(\theta) c(\varphi_2) & c(\theta) 
\end{pmatrix},
\end{eqnarray}
where $s$ and $c$ stand for sine and cosine.
It is assumed that the initiation time of a crack in a grain depends on the maximum shear stress in the slip systems of that grain. Nickel based alloys form a face centered cubic (FCC) lattice structure, therefore each grain has 4 slip planes with 3 different slip directions. To calculate the ratio of the externally applied strain and the shear stress in the slip systems of a grain, the anisotropic stress must first be calculated. Therefore the anisotopic stiffness tensor $C \in \mathbb{R}^{3 \times 3 \times 3 \times 3}$ must be rotated four times with the rotational matrix $U(\varphi_1, \theta, \varphi_2)$. 
\begin{figure}
\centering
\begin{subfigure}{0.6\textwidth}
    \includegraphics[width=\textwidth]{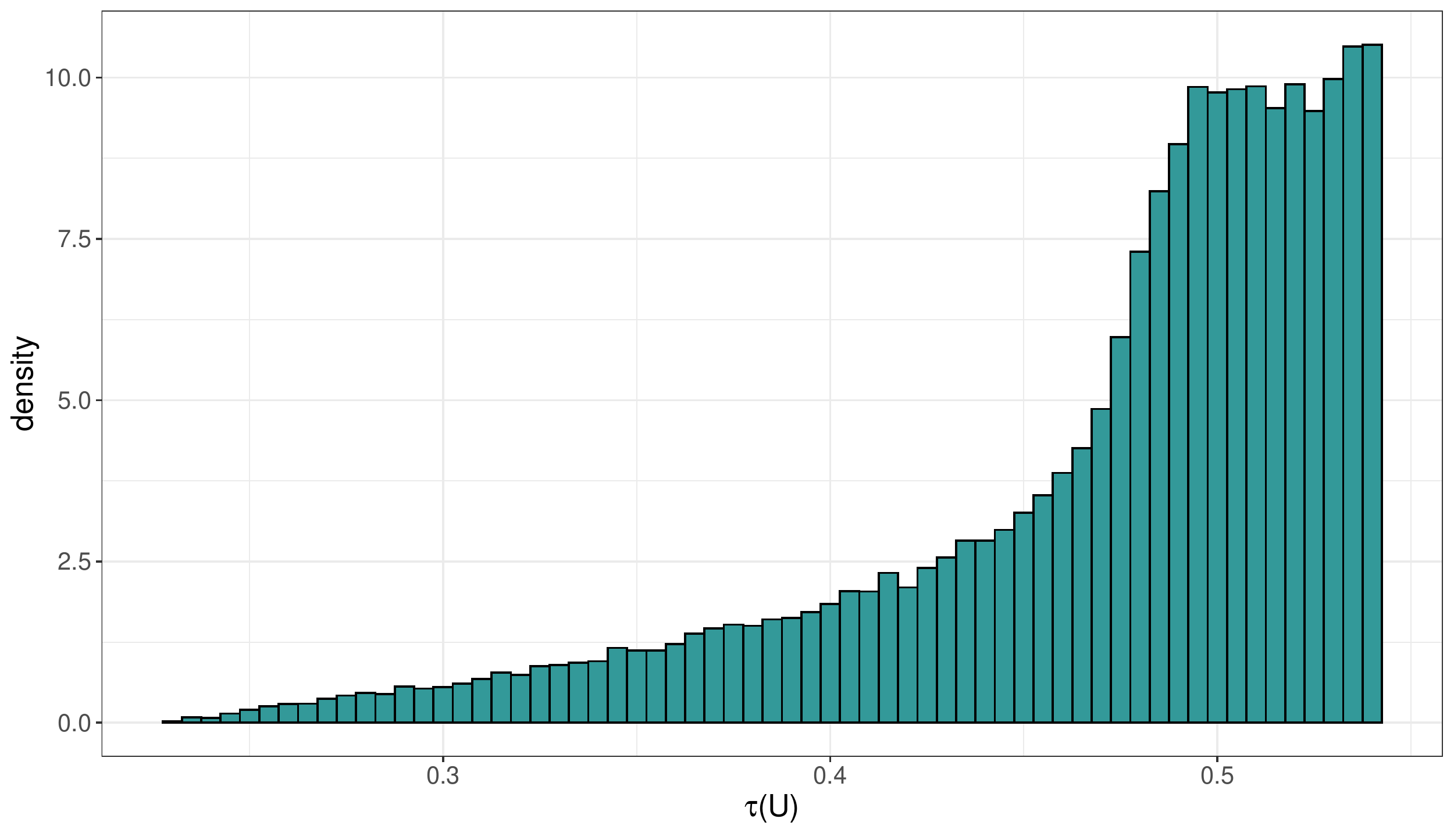}
    \caption{Histogram of the maximum shear stress in the slip systems of INC738 LC. 100000 random oriented grains under uniaxial load state.}
    \label{fig:hist_max_sf}
\end{subfigure}
\hfill
\begin{subfigure}{0.3\textwidth}
    \includegraphics[width=\textwidth]{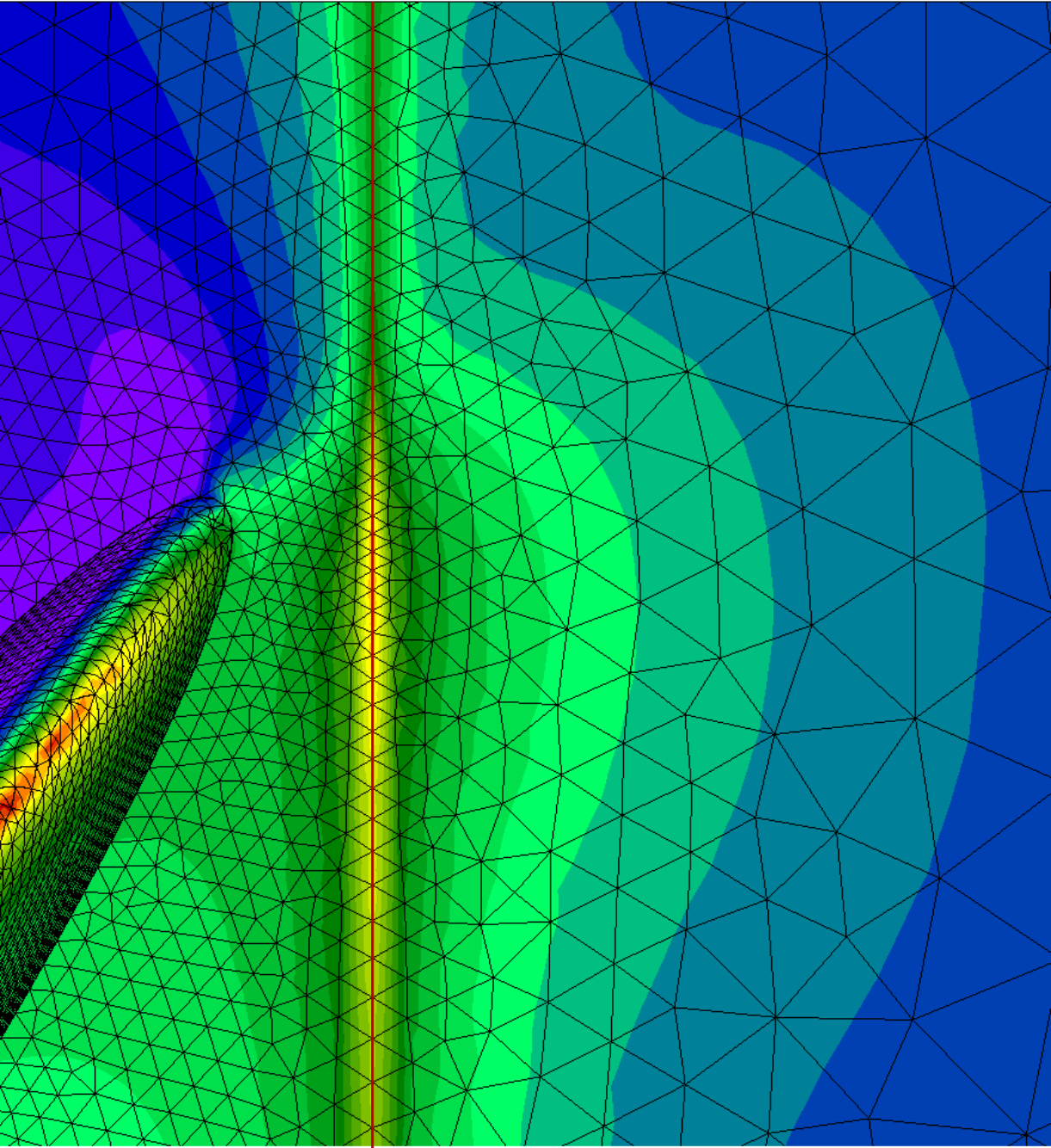}
     \caption{Geometry for the FEM simulations. Left grain cracked in the middle there is the grain boundary.}
    \label{fig:geo}
\end{subfigure}
\caption{Histogram of the maximum shear stress and closeup of the geometry in the FEM-simulations}\label{fig:SF_GEO}
\end{figure}
\vskip-.6cm
\begin{eqnarray}
C(U)_{ijkl} = \sum_{p,q,r,s}U_{ip} U_{jq} U_{kr} U_{ls} C_{pqrs}
\end{eqnarray}
To compute the anisotropic stress for a given strain tensor in each grain Hooke's law is used. 
\begin{eqnarray}
\sigma(U)^{\textup{ani}}_{ij} = \sum_{k,l} C(U)_{ijkl}  \varepsilon_{kl} \, .
\end{eqnarray}
With the anisotropic stress the resulting shear stress in the slip systems is calculated with the help of the stress tensor
\begin{eqnarray}
\tau_{ij}(U) = Un_{i} \cdot \sigma(U)^{\textup{ani}} \cdot Us_{ij} ,
\end{eqnarray}
where $s_{ij}$ is the direction and $n_i$ the normal of the slipsystem, gained with the orientation of the grain. For a rotation of the grain or elementary cell, the slip systems also rotates. For this purpose, the plane and directional vectors are rotated with the rotation matrix U mentioned in equation (\ref{rot_m}). Crack initiation is expected in the most heavily loaded slip system, therefore only the slip system with the maximum shear stress is considered in a grain.
\begin{eqnarray}
\tau(U) = \max_{ij} |\tau_{ij}(U)| \, .
\end{eqnarray}
So to have a ratio between maximum shear stress and the applied principal stress, the alternative Schmid factor
\vskip-.6cm
\begin{eqnarray}
m(U) = \dfrac{\tau(U)} {\sqrt{2/3} || \sigma_{iso} ||_F } ,
\end{eqnarray}
is used, which is introduced in \cite{MochPhD_2018}.
If we assume that the grains have no preferential orientation, the rotation matrices are distributed according to the isotropic measure, given by the Haar measure of the $SO(3)$ group of rotations \cite{LeonMasse2006}. 
The plot in figure \ref{fig:hist_max_sf} shows a histogram of $m(U)$ under uniaxial load. The lattice stiffness anisotropy is considered at 850°C for the alloy INC738 LC  \cite{EngelMaede_2019}.


\section{ANISOTROPIC FEM SIMULATIONS OF THE INFLUENCE OF CRACKS IN NEIGHBORING GRAINS}\label{sec:FEM}
A cracked grain changes the stress field in its surrounding and therefore the
shear stress acting on slip systems in neighboring grains changes as well. The change in shear stress 
leads to changes in the remaining life time\ after the crack is initiated. Our 
approach to determine this difference is to use FEM simulations. 
The geometry in the FEM simulation consists of two adjacent grains with a grain 
boundary parallel to the uni-axial loading direction. The left and the right 
grain have an orientation of the lattice structure each given by three Euler angels 
$(\varphi_1^i,\theta^i,\varphi_2^i), i \in \{r,l\}$. According to this, the 
stiffness tensor for the grains is given by $C(U^i)$, where $C(U^i)$ corresponds 
to the rotation matrix $U(\varphi^i_1, \theta^i, \varphi^i_2)$. The goal is to 
determine the new shear stress in each individual slip systems $\tau^{wc}_{ij}$  
resulting from the crack  as a function of the orientations of the two grains 
and the shear stress $\tau^{nc}_{ij}$ that was found in the slip system of the 
right grain before the crack

\begin{figure}[]
\centering
\begin{subfigure}{0.4\textwidth}
    \includegraphics[width=\textwidth]{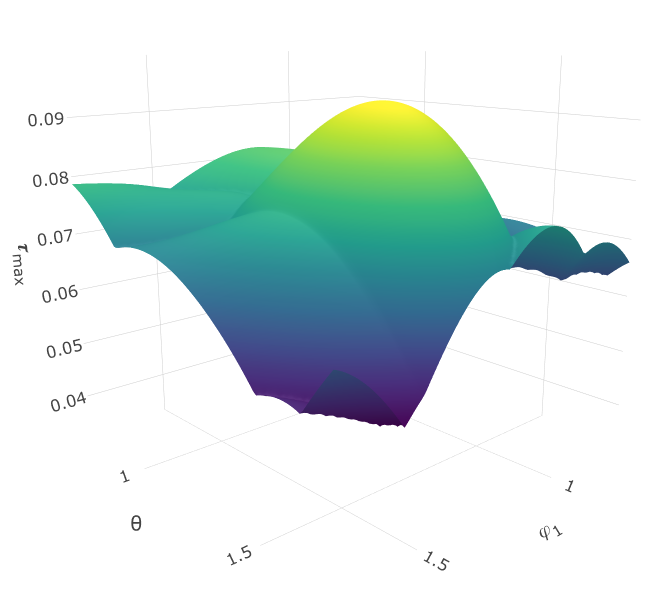}
    \caption{Maximal shear stress $\max_{ij}(\tau^{wc}_{ij}(U^r))$, fixed left grain.}
    \label{fig:tm_r}
\end{subfigure}
\hfill
\begin{subfigure}{0.4\textwidth}
    \includegraphics[width=\textwidth]{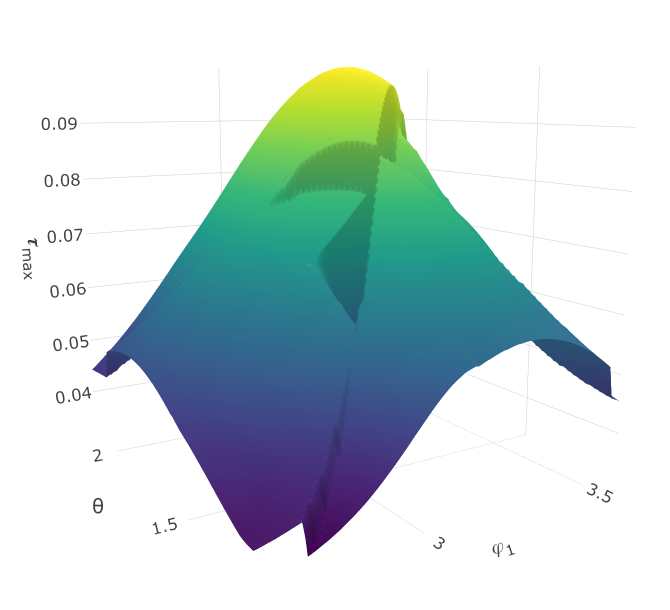}
     \caption{Maximal shear stress $\max_{ij}(\tau^{wc}_{ij}(U^r))$, fixed right grain.}
    \label{fig:tm_l}
\end{subfigure}
\caption{Maximum shear stress in the right grain with a cracked left grain. In the left plot, the orientation of the right grain is fix, the Euler angel $\varphi^l_1$ and $\theta^l$. In the right plot, the left grain is fixed while $\varphi_1^r$ and $\theta^r$ are changing}\label{fig:FEM_res}
\end{figure}

\vskip-.6 cm
\begin{eqnarray}
\tau^{wc}_{ij}(U^r) = a(\tau^{nc}_{ij}(U^r),\varphi^r_1, \theta^r, \varphi^r_2,\varphi^l_1, \theta^l, \varphi^l_2).
\end{eqnarray}
For this purpose, two simulations were evaluated for each pair of orientations. 
One with a half of a penny shaped crack in the left grain and one without a crack. The direction of the 
crack in the left grain depends on the slip  system with the maximum shear stress. 
For the geometry of the ellipsoidal crack, it is assumed that the whole grain is 
cracked. Accordingly, the crack length corresponds to the average diameter of a 
grain. The resulting shear stresses in the right grain are calculated by 
averaging over the volume of a half sphere centered on the grain boundary at the tip 
of the crack intersected with the volume of the right grain. 
\vskip-.6cm
\begin{eqnarray}
\label{eqn:int}
\bar{\tau}_{ij} = \dfrac{1}{|B|} \int_{x \in B} U^r n_{i} \cdot \sigma_{FEM}(x) \cdot U^rs_{ij} dx \, .
\end{eqnarray}
The integral in equation (\ref{eqn:int}) is calculated numerical with an Gaussian quadrature.

\begin{figure}[]
\centering
\begin{subfigure}{0.4\textwidth}
    \includegraphics[width=\textwidth]{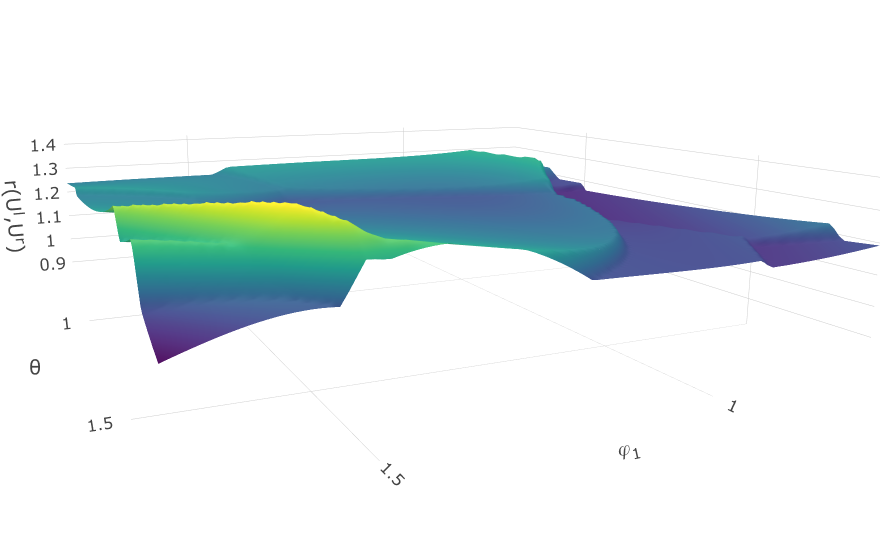}
    \caption{Ratio of the Maximal shear stress $r(U^r, U^l)$, fixed left grain}
    \label{fig:r_r}
\end{subfigure}
\hfill
\begin{subfigure}{0.4\textwidth}
    \includegraphics[width=\textwidth]{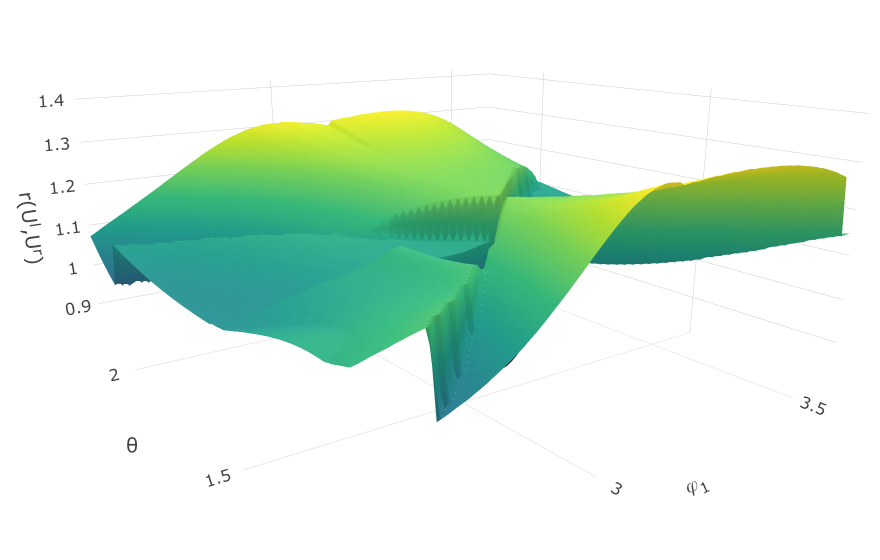}
    \caption{Ratio of the Maximal shear stress $r(U^r, U^l)$,fixed right grain.}
    \label{fig:r_l}
\end{subfigure}
\caption{Ratio of the maximum shear stress in the right grain $r(U^r, U^l)$. In the left plot, the orientation of the right grain is fix, the Euler angel $\varphi^l_1$ and $\theta^l$. In the right plot, the left grain is fixed while $\varphi_1^r$ and $\theta^r$ are changing}\label{fig:FEM_ratio}
\end{figure}

The toolchain for the simulations is as follows, the generation of the geometry and the pre-processing is done with R.
The grid is created using Gmsh. The FEM simulations are solved with CalculiX. As post-processor we use R code again.

\begin{wrapfigure}{hr}{0.55\textwidth}

  \begin{center}
\includegraphics[width=.9\linewidth]{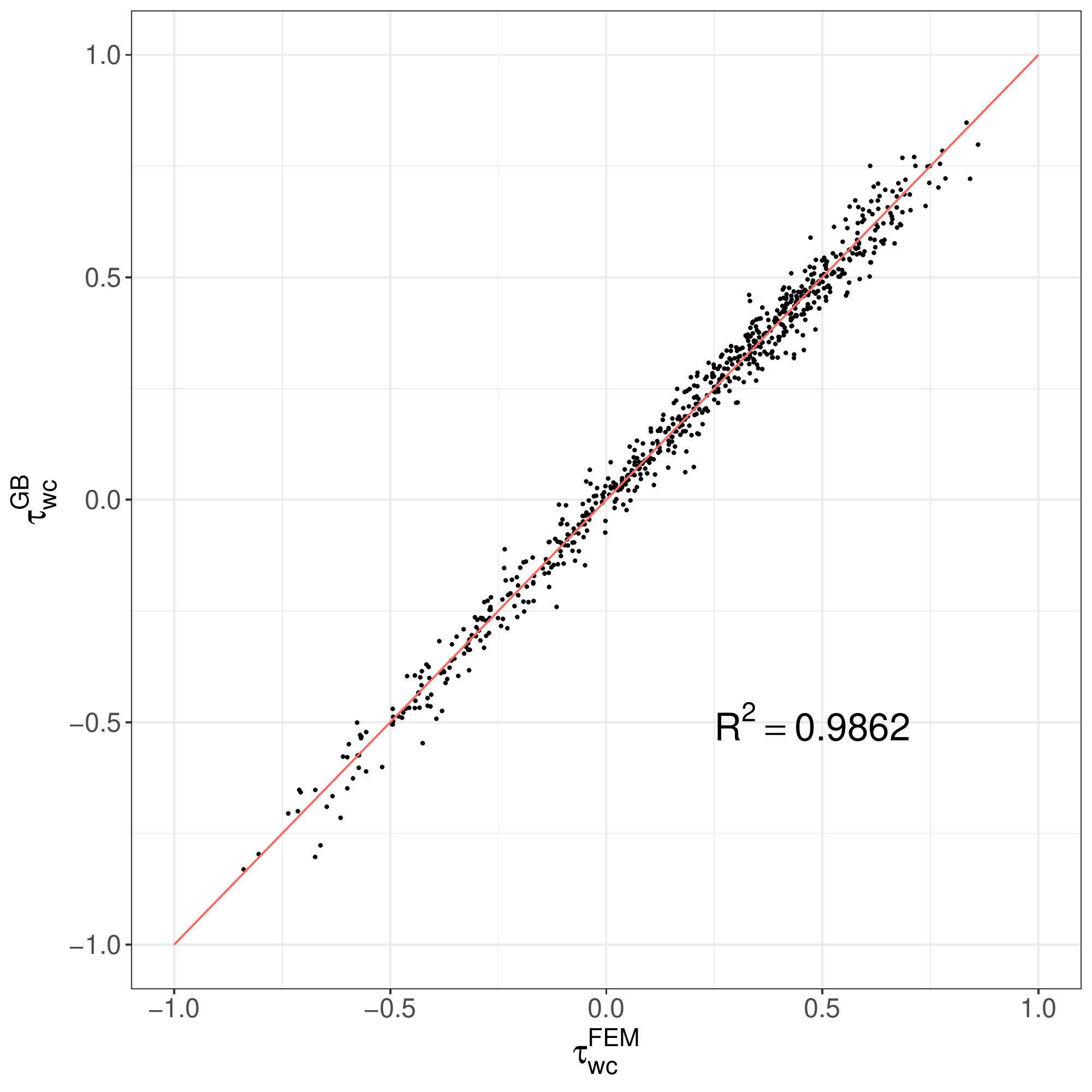}
  \end{center}
  \caption{Prediction of the shear stress vs results from the FEM-simulations.}
  \label{fig:train}
\end{wrapfigure}

The simulation is controlled and parallelized with R.
 The mesh consists of about $210000$ nodes and $130000$ tetrahedral elements of order $2$. 
One simulation takes about 15 minutes, with our current hardware setup we are able to run 40 simulations in parallel.

The results of the FEM simulations

for different orientations of the neighboring 
grains are shown in Figure~\ref{fig:FEM_res}.
As we can see in Figure \ref{fig:tm_r}, the maximum shear stress is continuously 
differentiable as a function of the orientations of the intact grain. 
Exceptions are the points where the slip system with the maximum shear stress 
changes, here the maximum shear stress is not differentiable. If the left grain is 
rotated, the slip system with the maximum shear stress changes as well, therefore the crack direction changes.

This leads to sudden change of the stress field in the intact grain, therefore the resulting 
shear stress in the slip systems shows the same response (\ref{fig:tm_l}). This 
behavior has an impact on the ratio of the maximal shear stress with 
and 
without crack
\vskip-.6cm
\begin{eqnarray}
r(U^r, U^l) = \frac{\max_{ij}(\tau^{wc}_{ij}(U^r))}{\max_{ij}(\tau^{nc}_{ij}(U^r))}  \, 
\end{eqnarray}

as shown in Figure \ref{fig:FEM_ratio}. Since the course of the shear stress is almost identical for a cracked 
and non-cracked left grain depending on the orientation of the right grain, the 
ratio is constant in large areas (\ref{fig:r_r}). The regions where the ratio 
changes are those where the maximum stressed slip system changes. For a fixed right 
grain, the maximal shear stress is continuous for a non-cracked left grain, 
therefore in the ratio $r(U^r, U^l)$, the sudden changes in the  shear stress 
show up as well (\ref{fig:r_l}).

To model the shear stress after a crack occurred as a function of the Euler angles we build a dataset from $25000$
simulations with randomly orientated grains distributed according to the isotropic Haar measure.

With this database,
a machine learning model is trained.  The input for the model are the six Euler angles that encode the respective orientation of the two grains and the 
shear stress in a particular slip system of the right grain if the left grain is not cracked.

The gradient boosting machine then predicts the shear stress with a cracked left grain. So for each of the 12 slip
systems one separate model is trained. For the machine learning model we use the implementation within the \textit{gbm} package in R. For the loss function we take squared error. The 
detailed settings are as follows: for the maximal number of trees we chose $2000$, the training fraction is $0.5$ and the interaction depth is set to $6$. With these setup, the R squared value for the 12 slip systems is between $0.970$ and $0.987$. The 
predictions over actual values for one particular slip system are shown in figure~\ref{fig:train}.

\section{THE EPIDEMIOLOGICAL CRACK PERCOLATION MODEL}\label{sec:Percolation}
To model the polycrystalline rectangular surface of a cast metal with height 
$h$ and length $l$, we use a Voronoi tessellation with $n$ seeds. The seeds are independently sample from the uniform distribution in the plane. A grain with seed $g$ then consists of every point for which the Euclidean distance to the seed $g$ is smaller then distance to any other seed. Two grains are adjacent if the share a common boundary. We assume that the grains have no preferential orientation. Accordingly the orientation is again sampled according to Haar's measure \cite{LeonMasse2006}.

For a given strain amplitude $\varepsilon_a$,  the same procedure as in section \ref{sec:Schmid} can now be used to determine shear stresses and Schmid factors.
to gain the stress amplitude from a mixture of plastic and elastic deformation 
the inverse of the  Ramberg–Osgood equation (RO) 
\vskip-.6cm
\begin{eqnarray}
RO(\sigma_a)  = \dfrac{\sigma_a}{E} + K \left(\dfrac{\sigma_a}{E} \right)^{n_{RO}},
\end{eqnarray}
with material dependent parameters $E,K,n_{RO}$ is used. With the so gained stress amplitude $\sigma_a = RO^{-1}(\varepsilon_a)$ the shear stress in each slip system of each grain is calculated and transformed back into an strain amplitude. Now  the inverse of the Coffin-Manson Basquin equation (CMB)
\vskip-.6cm
\begin{eqnarray}
CMB(N_i)  =\dfrac{\sigma'_f}{E}(2N_i)^b + \varepsilon'_f\cdot (2N_i)^c,
\end{eqnarray}
is used do gain the initial crack times for each grain: 
\vskip-.6cm
\begin{eqnarray}\label{init_time}
N^g_i  =CMB^{-1}\left(RO\left(RO^{-1}(\varepsilon_a) \dfrac{m(U^g)}{\lambda}\right)\right).
\end{eqnarray}
In equation (\ref{init_time}) $\lambda = \mathbb{E}[m(U)]$  is used as an scaling factor. With the initial failure times, the grain that cracks first can now be determined as $\hat{g} = \arg \min_g(N_i^g)$. 
Aa a grain $\hat{g}$ cracks, the neighboring grains are infected. So for a adjacent grain $g$ the new shear stress in the slip systems are calculated with the gradient boosting machines from section~\ref{sec:FEM}. 
With the new shear stress the Schmid factors $\bar{m}^g$ are calculated. To calculated the new crack time of $g$, we use a Miners rule which take the consumed lifetime into account:    
\vskip-.6cm
\begin{eqnarray}
    \bar{N}^g  = N^{\hat{g}} + \left(1 - \dfrac{N^{\hat{g}}}{N(m^g)}\right) \cdot N(\bar{m}^g) \, . 
\end{eqnarray}

The failure criterion for the whole surface that based on a
stress intensity factor K. Therefore after each cracked grain we evaluate the resulting crack clusters.
All cracked grains that are adjacent belong to the same crack cluster.
Each crack cluster is evaluated by taking all possible pairs of two grains in the cluster. 
The resulting direction of the total crack from these two grains is the vector $\vec{t}$ between the two seeds of the corresponding Voronoi cells, the normal is just the normal in the surface. Our multi modal stress intensity factor is as follows:
\vskip-.6cm
\begin{eqnarray}
    K_I = \vec{n}^T \cdot \sigma_{iso} \cdot \vec{n} \cdot \sqrt{\pi \dfrac{\lVert \vec{t} \rVert}{2}} \, ,
     \\
    K_{II} = \dfrac{\vec{t}^T}{ \lVert \vec{t} \rVert} \cdot \sigma_{iso} \cdot \vec{n} \cdot \sqrt{\pi \dfrac{\lVert \vec{t} \rVert}{2}} \, ,
    \\
    K_{eq} = \sqrt{K_I^2 + K_{II}^2} \, .
\end{eqnarray}
If $K_{eq}$ exceeds a critical value $K_{crit}$, there is a material failure.
\section{FITTING AND EXPERIMENTAL VALIDATION}
\label{sec:Fitting}
\begin{figure}
\centering
\begin{subfigure}{0.45\textwidth}
    \includegraphics[width=\textwidth]{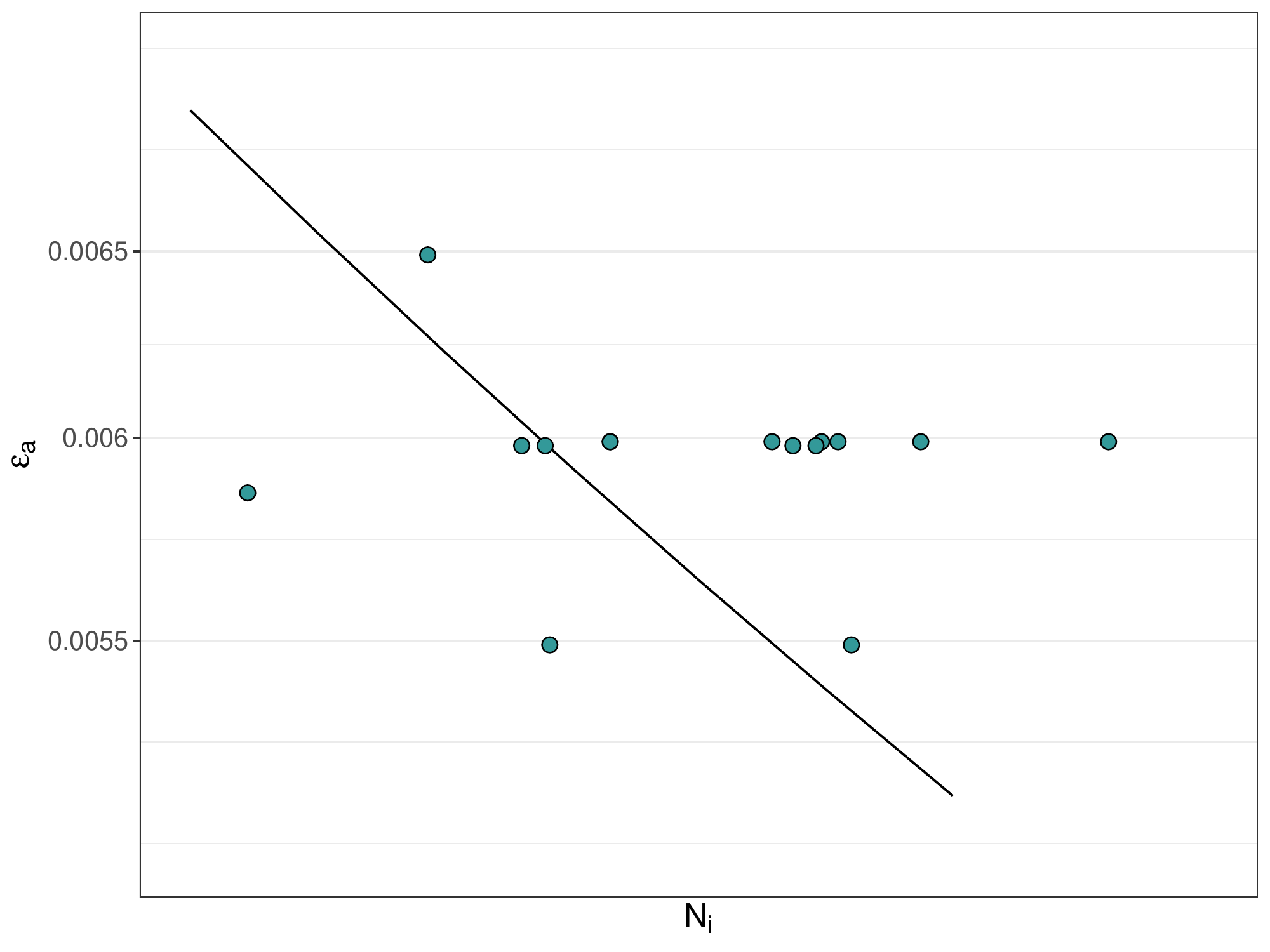}
    \caption{LCF experimental results with pre-fitted CMB relation }
    \label{fig:RENE80}
\end{subfigure}
\hfill
\begin{subfigure}{0.45\textwidth}
    \includegraphics[width=\textwidth]{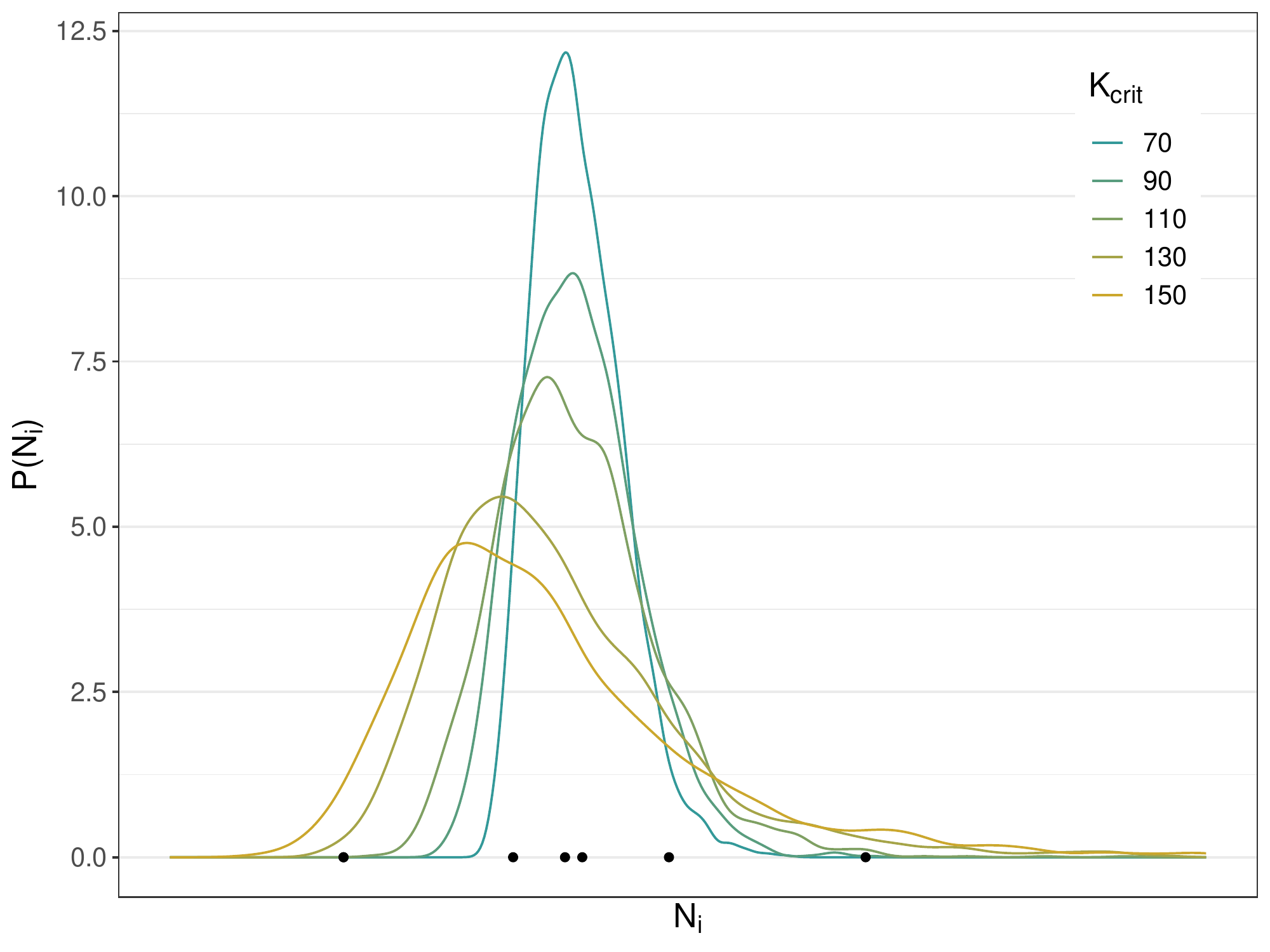}
     \caption{Probability of Failure for different values of $K_{crit}$}
    \label{fig:kcrit}
\end{subfigure}
\caption{Experimental LCF results (a) for different strain levels. Influence of the critical K-factor on PoF (b).  }\label{fig:dat_kcrit}
\end{figure}
To validate our percolation model, we use data from LCF experiments with RENE 80. The test specimens have
a gauge length of $18 mm$ and a radius of $7 mm$. The average diameter of the grains is $1.6mm$, therefore the surface contains of $198$ grains. The strain amplitude is between 0.55\% and 0.65\% for the specimens. The experiments were carried out at 850°C. As an failure criterion for the experiments we choose a drop in load of $5 \%$.

With these parameters, we perform 2500 simulations of the percolation model wit a pre-fitted CMB parameters per strain level. For comparison, we also run the simulation without the infection model. With the simulation data generated in this way, our model is now fitted to the experimental results. The parameters we fit are on the one hand a shift $s_{CMB}$ of the CMB which changes the position of the probability of failure (PoF)
\vskip-.6cm
\begin{eqnarray}
CMB'(\varepsilon_a) =CMB(\varepsilon_a) \cdot s_{CMB}  \, ,
\end{eqnarray}
on the other hand we adjust  the failure criterion $K_{crit}$, which mainly affects the scatter of the resulting PoF. The influence of different values of $K_{crit}$ on the PoF is shown in figure \ref{fig:kcrit}.  

\begin{figure}

\centering

    \includegraphics[width=0.59\textwidth]{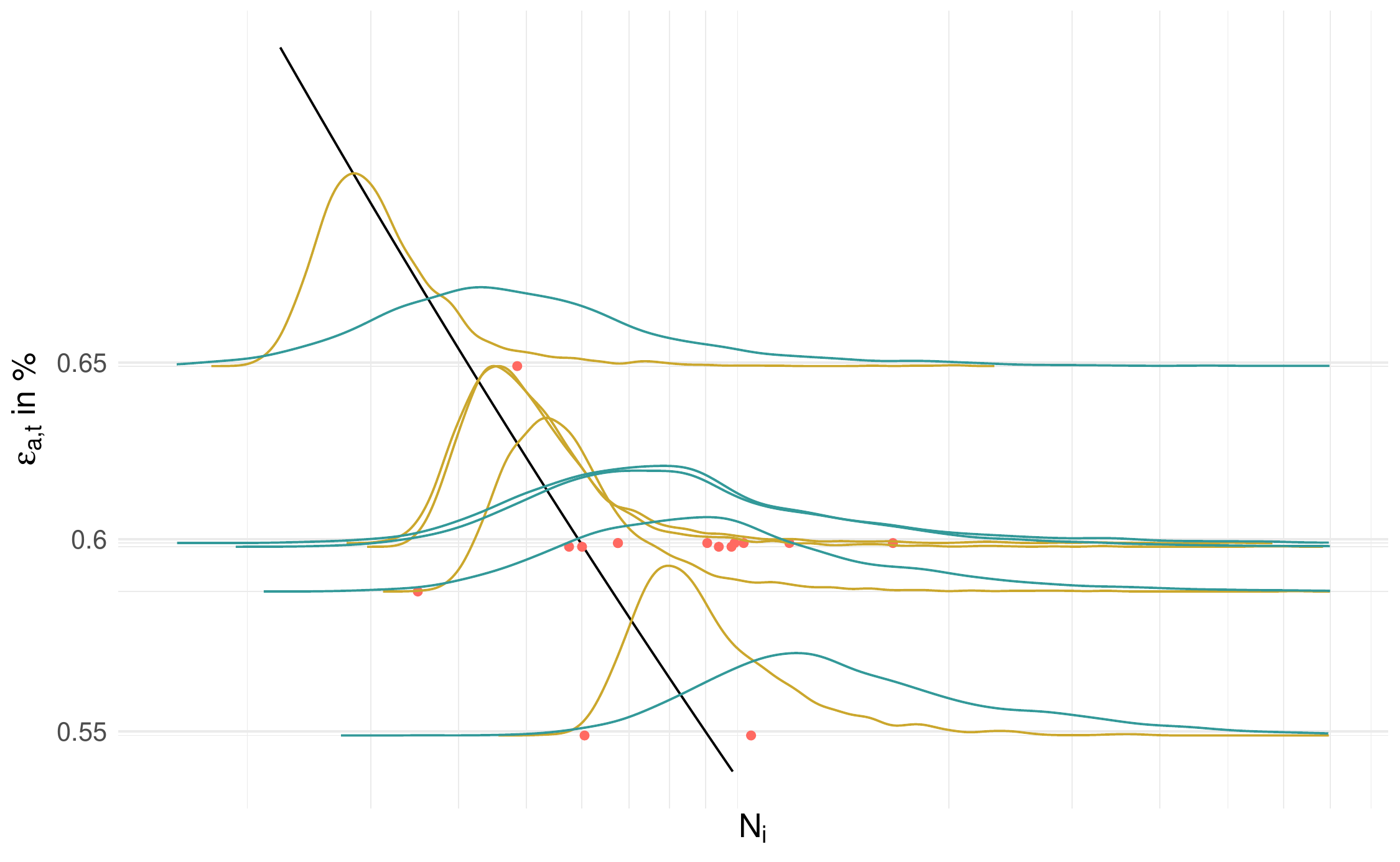}

\caption{Comparison of the probability of failure for the model with (yellow) and without infection (green).}\label{fig:percfit}
\end{figure}

The fitting of the maximum likelihood estimation is performed in R with the Nelder–Mead algorithm which is implemented in the package \textit{nloptr}. We compare the results from our epidemiological crack percolation model with the percolation model with out the infection function. As figure \ref{fig:percfit} shows, the interaction between the cracked grain and the surrounding grains reduce the the scatter in the resulting PoF significantly as compared to the percolation model without infection.

It turns out that both models, with and without infection, predict a statistical scatter bands in the physically correct order of magnitude. While the infection free model's prediction tends to be over-dispersed, this is removed by the model that includes infection. However, the data at strain amplitude $0.6\%$ seems to be right shifted in average, while the dispersion with is realistically modeled by the infection model. This, potentially, is due to a batch effect in the production of specimens or an experimental shift due to lab conditions as a second source of scatter. A final settlement of this issue would require further investigation and potentially more data as well. Also, including batch effects as random effects in the model could be a means for more accurate modeling.

\section{CONCLUSION}
\label{sec:Conclusion}

We presented a probabilistic model of low cycle fatigue (LCF) that models the scatter of fatigue experiments on the basis of an initial crack initiation phase and a subsequent crack percolation phase, where the neighboring grains of a cracked grain fail faster due to elevated stress levels. The statistical scatter in LCF cycles to crack initiation is explained from the effects of random grain structure, modeled by a Voronoi tessellation, and random orientation of the grains crystallographic axes. Here the latter effect accounts both for scatter due to anisotropic elasticity properties and also for the orientation of slip systems that enable plastic deformation.

We integrated a finite element simulation for the elevation of loads in grains with anisotropic elastic properties via a gradient boosting machine trained on the results of the finite element simulations. In this way, we are able to integrate elevated stress due to a crack in a neighboring grain in a probabilistic simulation of crack propagation. In this way, we pass from a simple model with independent cracks to an epidemiological percolation model of crack growth.

Finally, we compared our modeling approach with data from the Ni-based superalloy Rene80 and fitted the parameters of our model. We were able to show that the range of statistical scatter that is modeled is in the correct order of magnitude compared with the actually observed scatter and that the model can be reasonably fitted to the data for LCF experiments at high load amplitudes.   

In future research, we will further develop our model and account for effects due to deviations from uniform grain orientation at the surface of cast specimens due to dendrite growth of superalloy grains, in contrast to uniformly distributed orientations in bulk material. We also intend to further develop our model by including batch random effects and extend the range of validity from rather high load amplitudes to the high cycle fatigue regime.  
\printbibliography

 \vspace{6pt}
 \noindent\textbf{Acknowledgements.\space}{\textrm{The investigations were carried out as part of the joint project Roboflex of AG Turbo and funded
by the Federal Ministry for Economic Affairs and Climate Action (BMWK) under the grant number
03EE5013N.}}
 \par

\end{document}